\documentclass[conference]{IEEEtran}
\usepackage{amssymb,stmaryrd,amsmath,amsfonts,rotating}
\usepackage{amsmath}
\usepackage{amsthm}
\usepackage{color}
\usepackage{amssymb}
\usepackage{bm}
\usepackage{mathtools}
\mathtoolsset{showonlyrefs}

\newcommand{\R}{\mathbb{R}}

\newcommand{\dg}{g}
\newcommand{\dl}{l}

\newcommand{\dr}{r}

\newtheorem{theorem}{Theorem}

\newtheorem{definition}{Definition}

\newtheorem{lemma}{Lemma}

\newcommand{\es}{\epsilon^{{\rm{Sha}}}}

\newcommand{\x}{\bm{x}}
\newcommand{\y}{\bm{y}}

\newcommand{\D}{\bm{D}}
\newcommand{\f}{\bm{f}}
\newcommand{\g}{\bm{g}}

\newcommand{\Xc}{\mathcal{X}}

\newcommand{\Fc}{\mathcal{F}}

\newcommand{\Z}{\mathbb{Z}}
\renewcommand{\d}{\mathrm{d}}

\ifCLASSINFOpdf
\else
\fi
\begin{document}
%
\title{ Spatially-Coupled MacKay-Neal Codes with No Bit Nodes of Degree Two Achieve the Capacity of BEC}
\author{
\IEEEauthorblockN{Takuya Okazaki and Kenta Kasai}
\IEEEauthorblockA{Department of Communications and 
Computer Engineering,\\
 Tokyo Institution of Technology\\
Email: \{osa,kenta\}@comm.ce.titech.ac.jp}

}
\maketitle

\begin{abstract}
Obata et al. proved that spatially-coupled (SC) MacKay-Neal (MN) codes achieve the capacity of BEC. 
However, the SC-MN codes codes have many variable nodes of degree two and have higher error floors.
In this paper, we prove that SC-MN codes with no variable nodes of degree two
achieve the capacity of BEC.
\end{abstract}
\IEEEpeerreviewmaketitle

\section{Introduction}
Felstr\"{o}m and Zigangirov introduced spatially-coupled (SC) codes defined by sparse parity check matrix. 
SC codes are based on constitution method for convolutional LDPC codes \cite{zigangirov99}. 
Lantmaier et al. confirmed that regular SC LDPC codes achieve MAP threshold of original LDPC block codes by BP decoding in at least certain accuracy \cite{lentmaier_II}. 
Kudekar et al. proved that SC codes achieve MAP threshold by BP decoding on binary erasure channel (BEC) \cite{5695130} and binary symmetric channel \cite{6589171}.

Kasai et al.~introduced SC MacKay-Neal (MN) codes, and showed that these codes with finite maximum degree achieve capacity of BEC by numerical experiment \cite{HSU_MN_IEICE}. 
Obata et al. proved $(\dl,2,2)$ SC-MN codes achieve capacity \cite{ISIT_OJKP}. 
It has been observed that $(\dl,2,2)$ SC-MN codes have many bit nodes of degree two. This leads to high error floors. 

In this paper, we deal with $(\dl,3,3)$ SC-MN codes whose bit node degree is greater than two. 
We prove the codes achieve the capacity of BEC. 
The codes achieve Shannon limit $\es=1-\frac{3}{\dl}$ for any $\dl\ge 3. $
\section{Background}
\subsection{MacKay-Neal Codes}
$(l,r,g)$ MN codes are multi-edge type (MET) LDPC codes defined by pair of multi-variables degree distributions $(\mu,\nu)$ listed below.
\begin{equation}
\begin{split}
\nu(\x;\epsilon) &= \frac{\dr}{\dl}x_1^{\dl}+\epsilon x_{2}^{\dg},\\
\mu(\x) &=x_1^{\dr}  x_2^{\dg}.\\
\end{split}
\end{equation}
In general, the recursion of density evolution of MET-LDPC codes on BEC is given by
 \begin{eqnarray}
y_{j}^{(t)}&=&1-\frac{\mu_{j}(\bm{1}-\x^{(t)};1-\epsilon)}{\mu_j(\bm{1};1)}, \quad
x_{j}^{(t+1)}=\frac{\nu_{j}(\y^{(t)};\epsilon)}{\nu_{j}(\bm{1};1)},
 \end{eqnarray}
 where $x_j^{(t)}$ is probability of erasure message sent along edges of type $j$  at the $t$-th  decoding round. 
Therefore, density evolution of $(l,r,g)$ MN codes is
 \begin{align}
&    \x^{(t+1)} = \f\bigl(\g(\x^{(t)});\epsilon\bigr),\label{104340_21Jan14}\\
&\f(\x;\epsilon)= (x_{1}^{\dl-1},\epsilon x_{2}^{\dg-1} ),\\
&\g(\x)=(1-(1-x_{1})^{\dr-1} (1-x_{2})^{\dg},  1-(1-x_{1})^{\dr}(1-x_{2})^{\dg -1}  ).
\end{align}
\subsection{Spatially-Coupled MacKay-Neal Codes}
 SC-MN codes of coupling number $L$ and of coupling width $w$ are defined by the Tanner graph constructed by the following process. 
First,  at each section $i\in \Z$, place $rM/l$ bit nodes of type 1 and $M$ bits nodes of type 2. 
Bit nodes of type 1 and 2 are of degree $\dl$ and $\dg$, respectively. 
Next,  at each section $i \in \Z$, place $M$ check nodes of degree $\dr+\dg$. 
Then, connect edges uniformly at random so that 
bit nodes of type 1 at section $i$ are connected with check nodes at each section $i\in [i,\dotsc, i+w-1]$ with $\dr M/w$ edges, and
bit nodes of type 2 at section $i$ are connected with check nodes at each section $i\in [i,\dotsc, i+w-1]$ with $\dg M/w$ edges. 
Bits at section $i\notin [0,L-1])$ are shorten. 
Bits of type 1 and 2 at section $i\in[0,L-1]$ are punctured and transmitted, respectively. 
Rate of SC-MN codes $R^{MN}$ is given by
\begin{eqnarray*}
R^{\rm{MN}}&=&\frac{\dr}{\dl}+\frac{1+w-2\sum_{i=0}^{w}(1-(\frac{i}{w})^{\dr+\dg})}{L} 
=\frac{\dr}{\dl}\quad (L\to\infty).
\end{eqnarray*}
\subsection{Vector Admissible System and Potential Function}
In this section, we define vector admissible systems and potential functions. 
\begin{definition}
\label{def:vas}
{ 
Define $\Xc\triangleq[0,1]^d$, and   $F : \Xc \times [0, 1]\to \mathbb{R}$ and $G : \Xc \to \mathbb{R}$ as functionals satisfying $G(\bm{0})=0$. Let $\D$ be a $d \times d$ positive diagonal matrix. Consider a general recursion defined by
  \begin{equation}
    \x^{(t+1)}=\f(\g(\x^{(t)});\epsilon),
    \label{eq:1}
  \end{equation}
where $\f : \Xc \times [0,1] \to \Xc$ and $\g : \Xc \to \Xc$ are defined by  $F'(\x;\epsilon) = \f(\x;\epsilon) \D$ and $G'(\x) = \g(\x) \D$, 
where $F'(\x;\epsilon)\triangleq (\frac{\partial F(\x)}{\partial x_1},\dotsc,\frac{\partial F(\x)}{\partial x_n})$. 
Then the pair $(\f,\g)$ defines a vector admissible system if
 \begin{enumerate}
\item $\f,\g$ are twice continuously differentiable,
\item $\f(\x;\epsilon)$ and $\g(\x)$ are non-decreasing in  $\x$ and $\epsilon$ with respect to $\preceq$ \footnote{We say $\x \preceq \y$ if $x_i \leq y_i$ for all $1 \le i \le d$},
\item $\f(\g(\bm{0});\epsilon)=\bm{0}$ and $F(\g(\bm{0});\epsilon)=0$.
\end{enumerate}
We say  $\x$ is a fixed point if $\x=\f(\g(\x);\epsilon)$.
}
\end{definition}
It can be seen that the density evolution  $(\f,\g)$ of $(\dl,\dr,\dg)$ MN codes given in \eqref{104340_21Jan14} is a vector admissible system by choosing
 $F\bigl(\x;\epsilon\bigr), G(x)$ and $\D$ as below, since this system $(\f,\g)$ satisfies the condition in Definition \ref{def:vas}. 
 \begin{align*}
F(\x;\epsilon)&=\frac{\dr}{ \dl}x_{1}^{\dl}+\epsilon x_{2}^{\dg	},   \\
G(\x)&=\dr x_1+\dg x_2+(1-x_1)^{\dr}(1-x_2)^{\dg}-1,   \\
\D&=
\begin{pmatrix}
\dr &0\\
0&\dg
\end{pmatrix}. 
\end{align*}

\begin{definition}[{\cite[Def.~2]{simple_proof_BEC_itw}}]
  \label{def:potential}
 We define the potential function $U(\x;\epsilon)$ of a vector admissible system $(\f ,\g)$  by
 \begin{align}
  \label{eq:U}
  U(\x ;\epsilon) \triangleq \g(\x)\D\x^T-G(\x)-F(\g(\x);\epsilon).
 \end{align}
\end{definition}

The potential function $U(x_1,x_2,\epsilon)$ of $(l,r,g)$ MN codes is given by
\begin{align}
U(x_1,x_2,\epsilon)&=1-\epsilon \bigl((1-(1-x_1)^{\dr})(1-x_2)^{\dg-1}\bigr)^{\dg}\label{114201_21Jan14}\\
&-\frac{\dr}{\dl} \bigl(1-(1-x_1)^{\dr-1}(1-x_2)^{\dg}\bigr)^{\dl}\\
&-(1-x_1)^{\dr}(1-x_2)^{\dg}\Bigl(1+\frac{\dr x_1}{1-x_1}+\frac{\dg x_2}{1-x_2}\Bigr).
\end{align}

\begin{definition}[{\cite[Def. 7]{simple_proof_BEC_itw}}]
\label{def:PotentialThreshold}
{
 Let $\Fc(\epsilon) \triangleq \{\x \in \Xc\setminus\{\bm{0}\} \mid
\x=\f(\g(\x);\epsilon)\}$ be a set of non-zero fixed points for $\epsilon \in [0,1]$. The potential threshold $\epsilon^{*}$ is defined by
  \begin{equation*}
    \label{eq:3}
    \epsilon^{*} \triangleq \sup \{ \epsilon \in [0,1] \mid \min\nolimits_{\x \in \mathcal{F}(\epsilon)} U(\x;\epsilon) > 0 \}.
  \end{equation*}
Let $\epsilon_s^*$ be threshold of uncoupled system defined in \cite[Def. 6]{simple_proof_BEC_itw}. 
For $\epsilon$ such that $\epsilon_s^* < \epsilon < \epsilon^*$, we define energy gap $\Delta E(\epsilon)$ as
  $$\Delta E(\epsilon) \triangleq \max_{\epsilon'\in [\epsilon,1]} \inf_{ \x\in \mathcal{F}(\epsilon')} U(\x;\epsilon').$$
 }
\end{definition}

We define the SC system of a vector admissible system. 
\begin{definition}[{\cite[Def.~9]{simple_proof_BEC_itw}}]
\label{def:SpatiallyCoupled}
For  a  vector admissible system $(\f,\g)$, we define the SC system of coupling number $L$ and coupling width $w$ as 
 \begin{align}
\label{eq:4}
  \x^{(t+1)}_i &=
  \frac{1}{w}\sum_{k=0}^{w-1}\f\Bigg(\frac{1}{w}\sum_{j=0}^{w-1}\g(\x^{(t)}_{i+j-k});\epsilon_{i-k}\Bigg),\\
 \epsilon_{i} &= 
\begin{cases}
\epsilon,& i\in    \{0,\dotsc,L-1\},\\
 0,      & i\notin \{0,\dotsc,L-1\}.
\end{cases}
\end{align}
\end{definition}
If we define $(\f,\g)$ as the density evolution for $(\dl,\dr,\dg)$ MN codes in \eqref{104340_21Jan14}, the SC system gives
the density evolution of SC-MN codes with coupling number $L$ and coupling width $w$. 

Next theorem states that if $\epsilon < \epsilon^*$ then fixed points of SC vector system converge towards $\bold{0}$ for sufficiently large $w$.
\begin{theorem}[{\cite[Thm.~1]{simple_proof_BEC_itw}}]
\label{theorem:sc_theorem}
Consider the constant $K_{\f,\g}$ defined in \rm{\cite[Lem.~11]{simple_proof_BEC_itw}}. 
This constant value depends only on $(\f,\g)$. 
If $\epsilon < \epsilon^*$ and $w > (d K_{\f,\g})/( 2\Delta E(\epsilon))$, then the SC system of $(\f, \g)$ with coupling number $L$ and coupling width $w$ has a unique fixed point $\bold{0}$.
\end{theorem}
We will show that the potential threshold $\epsilon^*$ of $(\dl,\dr=3,\dg=3)$ MN codes is $1-R^{\rm{MN}}=1-3/\dl$ for any $\dl\ge 3$. 
This is sufficient to show that $(\dl,3,3)$ SC-MN codes with sufficiently large $w$ and $L$ achieve the capacity of BEC under BP decoding. 

\section{Proof of Achieving Capacity}
In this section, we calculate the potential threshold $\epsilon^*$ of $(\dl,\dr=3,\dg=3)$ MN codes. 
To this end, we first investigate the set of fixed points $\mathcal{F}(\epsilon)$. 

The density evolution recursion  in \eqref{104340_21Jan14}
can be rewritten as
\begin{align*}
x_{1}^{(t+1)}&=(1-(1-x^{(t)}_{1})^{\dr-1}(1-x^{(t)}_{2})^{\dg})^{\dl-1},\\
x_{2}^{(t+1)}&=\epsilon(1-(1-x^{(t)}_{1})^{\dr}(1-x^{(t)}_{2})^{\dg-1})^{\dg-1}.
\end{align*}
Fixed points $(x_1,x_2;\epsilon)$ of density evolution with $x_1=0$ and $x_1=1$ are $(0,0;\epsilon)$ and $(1,\epsilon;\epsilon)$, respectively. 
We define these fixed points as trivial fixed points and all other fixed points as non-trivial fixed points. 
All non-trivial fixed points $(x_1,x_2(x_1);\epsilon(x_1))$ can be parametrically described  as
\begin{align}
x_2(x_{1})=& 1- \biggl(\frac{1-x_{1}^{\frac{1}{\dl-1}}}{(1-x_1)^{\dr-1}}\biggr)^{\frac{1}{\dg}}, \label{eq:mnfpx2}\\
\epsilon(x_{1})=& \frac{x_{2} (x_{1})}{\bigl(1-(1-x_{1})^{\dr}(1-x_{2}(x_1))^{\dg-1}\bigr)^{\dg-1}},\label{eq:mnfpep}
\end{align}
with $x_1 \in (0,1)$.

\begin{figure}[t]
\setlength{\unitlength}{1.0bp}%
 \begin{picture}(100,150)(0,0)
 \put(-20,-40){\includegraphics{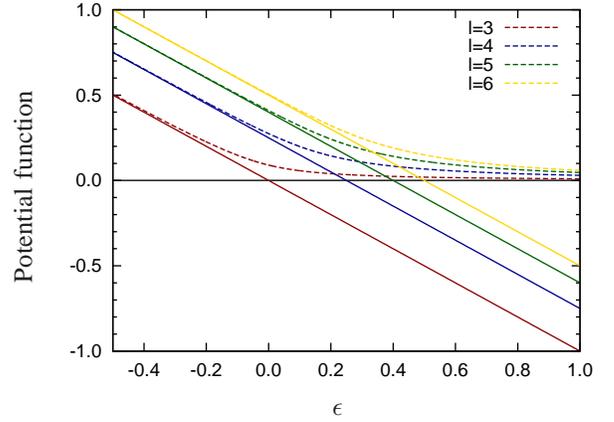}}
 \put(140,-00){$\epsilon$}
 \put(20,50){\rotatebox{90}{Potential function}}
 \end{picture}
\caption{ Potential function $U(\bm{1};\epsilon) $ and $U(\x(x_1);\epsilon(x_1))$ at the trivial fixed points (solid) and non-trivial fixed points (dashed) of $(l,3,3)$ MN codes for $l=3,\dotsc,6$}
 \label{fig:potential}
\end{figure}

Next, we shall investigate the value of the potential function at the fixed points. 
The value of the potential functions at trivial fixed point $(1,\epsilon,\epsilon)$ is respectively given by
\begin{eqnarray}
U(1,\epsilon,\epsilon)&=&1-\frac{\dr}{\dl}-\epsilon .   \label{eq:U1}
\end{eqnarray}
Figure \ref{fig:potential} draws the potential function of $(\dl,\dr,\dg)$ MN codes at fixed points $\x\in \mathcal{F}(\epsilon)$. 
It appears that  the potential function at non-trivial fixed points is always positive.
We will prove this. To be precise, the potential function of $(l,r,g)$ MN codes for non-trivial fixed points satisfies
 \begin{align}
U(x_1,x_2(x_1),\epsilon (x_1)) > 0  \ \  \text{ for } \ \   x_1 \in (0,1).\label{153641_18Jan14}
 \end{align}

Our strategy of proof is as follows. First change the representation of \eqref{153641_18Jan14} into a polynomial form by changing variables a few times. 
Then apply Sturm's theorem for smaller $\dl$ and bound the polynomial for larger $\dl$. 

 We define $U(z):=U(x_1,x_2(x_1),\epsilon (x_1)) |_{x_1=z^{\dl-1}}$. 
Obviously, to prove \eqref{153641_18Jan14}, it is sufficient to show  $U(z) > 0$ for $z\in (0,1)$. 
\begin{align}
  \label{eq:Uz}
 U(z)
 &=-\frac{3z^{\dl}}{\dl}+(1-z)(1-4z^{\dl-1})\\
 & +(1-z)^{1/3}(1-z^{\dl-1})^{-2/3} -2(1-z)^{2/3}(1-z^{\dl-1})^{5/3}.
\end{align} 
We use next lemma to eliminate fractional power in $U(z)$. The proof is given in Section \ref{154841_18Jan14}. 
 \begin{lemma}\label{lemma:Huz}
Define $H(u,z)$ as follows. 
\begin{align}
    H(u,z)&=\bigl(u+\frac{3z^{\dl}}{\dl}-(1-z)(1-4z^{\dl-1})\bigr)^3\\
    & +6(1-z)(1-z^{\dl-1})\bigl(u+\frac{3z^\dl}{\dl}-(1-z)(1-4z^{\dl-1})\bigr) \\
   & -(1-z)(1-z^{\dl-1})^{-2}+8(1-z)^{2}(1-z^{\dl-1})^5. 
\end{align}
Then, $H(0,z) < 0 $ for $z\in(0,1)$ implies $U(z) > 0$ for $z\in(0,1)$. 
 \end{lemma}
Define $\displaystyle I(z):=\frac{\dl^3(1-z^{\dl-1})^2}{(1-z)z^2}H(0,z)$. 
Obviously, to prove $H(0,z)<0$ for $z\in(0,1)$, it is sufficient to prove $I(z)<0$ for $z\in(0,1)$. 
We see that $I(z)$ for $\dl\ge 3$ is a polynomial as follows.
\begin{align}
  I(z)=&-l^3+27\sum_{i=0}^{l-2}[z^{3l-2+i}(1 - z^{l-1})]\\
  &-27l^2z^{-2 + 2l}(1 - 4z^{l-1}) (1 - z^{l-1})^2 \\
  &- 9lz^{-4 + l}(1 - z^{l-1})^2\bigl\{(-3 + z)z^2\\
  &+ 16(-1 + z)z^{2l}- 8(-1 + z)z^{1 + l}\bigr\} \\
  &- l^3(1 - z)z^{-9 + l}\bigl\{8z^{6l} - 56z^{1 + 5l} + 2z^6 (3 + 7z)\\
&  + 8z^{2 + 4l}(13 + 8z)- 8z^{3 + 3l}(13 + 22z)\\
&  + 4z^{4 + 2l}(21 + 43z) -   z^{5 + l}(41 + 73z)\bigl\}.
 \label{eq:I}
\end{align}
We prove $I(z) < 0$ for $3 \le l < 165$ and $\dl \ge 165$ in the following lemmas. 
The proofs are given in Section \ref{sec:145948_18Jan14} and Section \ref{sec:154724_18Jan14}, respectively.  
\begin{lemma}\label{lemma:I}
For $3 \le l <165$, $I(z) < 0$ for $z\in(0,1)$.
\end{lemma}
\begin{lemma}\label{lemma:II}
For $\dl\ge 165$, $I(z) < 0$ for $z\in(0,1)$.
\end{lemma}
\begin{theorem}
For any $\dl\ge 3$ and $\epsilon < \es=1-\frac{3}{\dl}$, the unique fixed point of density evolution of $(\dl,3,3)$ SC-MN codes of coupling number $L$ and coupling width $w$ is $\bold{0}$ for sufficiently large $w$ and $L$.
 \end{theorem}
Proof:
From  \eqref{153641_18Jan14}, potential function for non-trivial fixed points is always positive. Therefore, from Definition \ref{def:PotentialThreshold} and potential function for trivial fixed point \eqref{eq:U1}, $\epsilon^{*}=1-\frac{\dr}{\dl}=\es$.
From Theorem \ref{theorem:sc_theorem}, for $\epsilon <\es$, the unique fixed point of density evolution for  $(l,3,3)$ SC-MN codes is $\bold{0}$.
\qed

The case with $\dl=3$ implies rate one codes over BEC($0$). 
Some might think this is not interesting. 
Nevertheless, we included the case with $\dl=3$ for comprehensiveness.

\section{Proof of lemmas}
\subsection{Proof of Lemma \ref{lemma:Huz}}
\label{154841_18Jan14}
Partial derivative of $H(u,z)$ with respect to $u$ gives
  \begin{equation}
   \label{eq:dH}
   \begin{split}
   \frac{\partial H(u,z)}{\partial u}=&3\bigr(u+\frac{3z^\dl}{\dl}-(1-z)(1-4z^{\dl-1})\bigl)^2 \\
   & +6(1-z)(1-z^{\dl-1}) \ge 0 .
   \end{split}
  \end{equation}
Substituting $u=U(z)$ into $H(u,z)$ gives
  \begin{align}
    H&(U(z),z)\\
	=&\bigl((1-z)^{1/3}(1-z^{l-1})^{-2/3}-2(1-z)^{2/3}(1-z^{l-1})^{5/3}\bigr)^3\\
    &+6(1-z)(1-z^{l-1})\bigr\{(1-z)^{1/3}(1-z^{l-1})^{-2/3}\\
    &-2(1-z)^{2/3}(1-z^{l-1})^{5/3}\bigl\}\\
    &-(1-z)(1-z^{l-1})^{-2}+8(1-z)^{2}(1-z^{l-1})^{5}\\
    =&(1-z)(1-z^{l-1})^{-2}-8(1-z)^{2}(1-z^{l-1})^{5}\\
    &-6(1-z)(1-z^{l-1})\bigl\{(1-z)^{1/3}(1-z^{l-1})^{-2/3}\\
    &-2(1-z)^{2/3}(1-z^{l-1})^{5/3}\bigr\}\\
    &+6(1-z)(1-z^{l-1})\bigl\{(1-z)^{1/3}(1-z^{l-1})^{-2/3}\\
    &-2(1-z)^{2/3}(1-z^{l-1})^{5/3}\bigr\}\\
    &-(1-z)(1-z^{l-1})^{-2}+8(1-z)^{2}(1-z^{l-1})^{5}\\
    =&0.   \label{eq:HUz}
  \end{align}
 From \eqref{eq:dH}, $H(u,z)$ monotonically increasing with respect to  $u$. 
From \eqref{eq:HUz}, $(u,z)=(U(z),z)$ is a root of $H(u,z)=0$. 
Therefore $H(0,z) < 0 $ for $z\in(0,1)$ implies $U(z) > 0$ for $z\in(0,1)$. \qed

\subsection{Proof of Lemma \ref{lemma:I}}
\label{sec:145948_18Jan14}
From $I(0)=-l^3$ and $I(1)=-l^3$, we see that $z=0,1$ are not multiple roots of equation $I(z)=0$.  
Let $I_1(z),\dotsc,I_m(z)$ be Sturm sequences of $I(x)$. 
Let $V(z)$ be the number of sign changes in the sequence. 
Table \ref{tb:sturml} lists 
sign changes of Sturm sequence $I_1(z),\dotsc,I_m(z)$ of $I(x)$ in \eqref{eq:I}  for $\dl=3,\dotsc,11$. 
$V(z)$ is the number of sign changes in the sequence. 
We see that $V(0)=V(1)$. We observed that $V(0)=V(1)$ for $\dl< 165$ but not listed all due to the space limit. 
From Theorem \ref{theorem:sturm}, this implies that the number of distinct roots of equation $I(z)=0$ in $(0, 1]$ is $V(0)-V(1)=0$. 
Therefore, $I(z)<0, z\in(0,1)$ for $3, \dotsc, 164$. \qed
 \begin{table*}[t]
\caption{Sign changes of Sturm sequence $I_1(z),\dotsc,I_m(z)$ of $I(x)$ in (\ref{eq:I}) for $\dl=3,\dotsc,11$. $V(z)$ is the number of sign changes in the sequence.}
{
  \begin{tabular}{c|c|c|c|c@{\hspace{-0.1mm}}c@{\hspace{-0.1mm}}c@{\hspace{-0.1mm}}c@{\hspace{-0.1mm}}c@{\hspace{-0.1mm}}c@{\hspace{-0.1mm}}c@{\hspace{-0.1mm}}c@{\hspace{-0.1mm}}c@{\hspace{-0.1mm}}c@{\hspace{-0.1mm}}c@{\hspace{-0.1mm}}c@{\hspace{-0.1mm}}c@{\hspace{-0.1mm}}c@{\hspace{-0.1mm}}c@{\hspace{-0.1mm}}c@{\hspace{-0.1mm}}c@{\hspace{-0.1mm}}c@{\hspace{-0.1mm}}c@{\hspace{-0.1mm}}c@{\hspace{-0.1mm}}c@{\hspace{-0.1mm}}c@{\hspace{-0.1mm}}c@{\hspace{-0.1mm}}c@{\hspace{-0.1mm}}c@{\hspace{-0.1mm}}c@{\hspace{-0.1mm}}c@{\hspace{-0.1mm}}c@{\hspace{-0.1mm}}c@{\hspace{-0.1mm}}c@{\hspace{-0.1mm}}c@{\hspace{-0.1mm}}c@{\hspace{-0.1mm}}c@{\hspace{-0.1mm}}c@{\hspace{-0.1mm}}c@{\hspace{-0.1mm}}c@{\hspace{-0.1mm}}c@{\hspace{-0.1mm}}c@{\hspace{-0.1mm}}c@{\hspace{-0.1mm}}c@{\hspace{-0.1mm}}c@{\hspace{-0.1mm}}c@{\hspace{-0.1mm}}c@{\hspace{-0.1mm}}c@{\hspace{-0.1mm}}c@{\hspace{-0.1mm}}c@{\hspace{-0.1mm}}c@{\hspace{-0.1mm}}c@{\hspace{-0.1mm}}c@{\hspace{-0.1mm}}c@{\hspace{-0.1mm}}c@{\hspace{-0.1mm}}c@{\hspace{-0.1mm}}c@{\hspace{-0.1mm}}c@{\hspace{-0.1mm}}c@{\hspace{-0.1mm}}c@{\hspace{-0.1mm}}c@{\hspace{-0.1mm}}c@{\hspace{-0.1mm}}c@{\hspace{-0.1mm}}c@{\hspace{-0.1mm}}c@{\hspace{-0.1mm}}c@{\hspace{-0.1mm}}c@{\hspace{-0.1mm}}c@{\hspace{-0.1mm}}c@{\hspace{-0.1mm}}c@{\hspace{-0.1mm}}c@{\hspace{-0.1mm}}c@{\hspace{-0.1mm}}c@{\hspace{-0.1mm}}c@{\hspace{-0.1mm}}c@{\hspace{-0.1mm}}c@{\hspace{-0.1mm}}c@{\hspace{-0.1mm}}c@{\hspace{-0.1mm}}c@{\hspace{-0.1mm}}c@{\hspace{-0.1mm}}c@{\hspace{-0.1mm}}c@{\hspace{-0.1mm}}c@{\hspace{-0.1mm}}c@{\hspace{-0.1mm}}c@{\hspace{-0.1mm}}c@{\hspace{-0.1mm}}c@{\hspace{-0.1mm}}c@{\hspace{-0.1mm}}c@{\hspace{-0.1mm}}c@{\hspace{-0.1mm}}c@{\hspace{-0.1mm}}c@{\hspace{-0.1mm}}c@{\hspace{-0.1mm}}c@{\hspace{-0.1mm}}c@{\hspace{-0.1mm}}c@{\hspace{-0.1mm}}c@{\hspace{-0.1mm}}c@{\hspace{-0.1mm}}c@{\hspace{-0.1mm}}c@{\hspace{-0.1mm}}c@{\hspace{-0.1mm}}c@{\hspace{-0.1mm}}c@{\hspace{-0.1mm}}c@{\hspace{-0.1mm}}c@{\hspace{-0.1mm}}c@{\hspace{-0.1mm}}c@{\hspace{-0.1mm}}c@{\hspace{-0.1mm}}c@{\hspace{-0.1mm}}c@{\hspace{-0.1mm}}c@{\hspace{-0.1mm}}c@{\hspace{-0.1mm}}c@{\hspace{-0.1mm}}c@{\hspace{-0.1mm}}c@{\hspace{-0.1mm}}c@{\hspace{-0.1mm}}c@{\hspace{-0.1mm}}c@{\hspace{-0.1mm}}c@{\hspace{-0.1mm}}c@{\hspace{-0.1mm}}c@{\hspace{-0.1mm}}c@{\hspace{-0.1mm}}c@{\hspace{-0.1mm}}c@{\hspace{-0.1mm}}c@{\hspace{-0.1mm}}c@{\hspace{-0.1mm}}c@{\hspace{-0.1mm}}c@{\hspace{-0.1mm}}c@{\hspace{-0.1mm}}c@{\hspace{-0.1mm}}c@{\hspace{-0.1mm}}c@{\hspace{-0.1mm}}c@{\hspace{-0.1mm}}}
$l$&$m$&$V(z)$&$z$&\multicolumn{80}{|c}{$\mathrm{sgn}[I_0(z)],\mathrm{sgn}[I_1(z)],\dotsc, \mathrm{sgn}[I_m(z)]\hspace{10cm}$}\\\hline
   $3$&13&5&$0$&$-$&$-$&$+$&$+$&$+$&$-$&$-$&$-$&$+$&$-$&$-$&$-$&$+$&$+$\\
        &&5&$1$&$-$&$-$&$+$&$+$&$+$&$+$&$+$&$-$&$-$&$+$&$-$&$-$&$-$&$+$\\\hline  
   $4$ &20  &10 & $0$&$-$&$-$&$+$&$-$&$-$&$+$&$-$&$-$&$-$&$+$&$+$&$-$&$+$&$+$&$+$&$-$&$-$&$+$&$-$&$-$&$-$ \\ 
   & &10 & $1$&$-$&$-$&$+$&$-$&$-$&$-$&$+$&$-$&$-$&$+$&$+$&$-$&$-$&$-$&$-$&$+$&$-$&$-$&$+$&$-$&$-$ \\\hline
   $5$ &27  &12 & $0$&$-$&$0$&$+$&$-$&$-$&$-$&$+$&$+$&$-$&$-$&$+$&$+$&$+$&$-$&$+$&$+$&$-$&$-$&$-$&$+$&$+$&$-$&$-$&$-$&$+$&$-$&$-$&$-$ \\ 
   & &12 & $1$&$-$&$-$&$+$&$+$&$-$&$-$&$-$&$+$&$+$&$-$&$+$&$+$&$+$&$+$&$-$&$-$&$+$&$+$&$+$&$-$&$-$&$+$&$+$&$-$&$-$&$+$&$+$&$-$\\\hline
   $6$ &33    &16&0&$-$&$0$&$+$&$-$&$-$&$-$&$+$&$+$&$-$&$+$&$+$&$+$&$-$&$-$&$+$&$+$&$-$&$-$&$+$&$+$&$+$&$-$&$+$&$+$&$-$&$-$&$+$&$+$&$-$&$-$&$+$&$-$&$-$&$-$\\
   & &16&1&$-$&$-$&$+$&$+$&$-$&$-$&$-$&$-$&$+$&$-$&$+$&$+$&$-$&$-$&$-$&$-$&$+$&$+$&$-$&$-$&$-$&$+$&$+$&$-$&$+$&$-$&$-$&$-$&$+$&$-$&$+$&$+$&$+$&$-$\\\hline
   $7$&39 & 18 & 0&$-$&$0$&$+$&$-$&$-$&$-$&$+$&$+$&$-$&$-$&$+$&$+$&$-$&$+$&$+$&$+$&$+$&$-$&$+$&$+$&$+$&$-$&$-$&$+$&$+$&$-$&$-$&$+$&$+$&$-$&$-$&$-$&$+$&$+$&$+$&$-$&$+$&$+$&$+$&$-$\\
   &&18&1&$-$&$-$&$+$&$+$&$-$&$-$&$-$&$+$&$+$&$-$&$-$&$+$&$-$&$+$&$+$&$+$&$+$&$+$&$-$&$-$&$-$&$+$&$+$&$-$&$-$&$+$&$+$&$-$&$-$&$+$&$+$&$-$&$+$&$+$&$+$&$-$&$-$&$+$&$+$&$-$\\\hline
   $8$&45&22&0&$-$&$0$&$+$&$-$&$-$&$-$&$+$&$+$&$+$&$-$&$+$&$-$&$-$&$-$&$+$&$-$&$-$&$-$&$+$&$+$&$-$&$-$&$+$&$+$&$-$&$-$&$-$&$+$&$+$&$-$&$-$&$+$&$-$&$+$&$+$&$+$&$+$&$-$&$-$&$+$&$-$&$-$&$-$&$+$&$+$&$-$\\
   &&22&1&$-$&$-$&$+$&$+$&$-$&$-$&$-$&$+$&$+$&$+$&$-$&$+$&$-$&$-$&$+$&$-$&$-$&$-$&$-$&$-$&$+$&$+$&$-$&$-$&$+$&$+$&$-$&$-$&$-$&$+$&$+$&$-$&$+$&$-$&$-$&$-$&$+$&$-$&$-$&$+$&$+$&$+$&$-$&$+$&$+$&$-$\\\hline
   $9$&51&24&0&$-$&$0$&$+$&$-$&$-$&$-$&$+$&$+$&$-$&$-$&$+$&$+$&$-$&$-$&$+$&$-$&$+$&$+$&$+$&$+$&$+$&$-$&$+$&$+$&$+$&$-$&$-$&$+$&$+$&$-$&$-$&$-$&$+$&$-$&$-$&$+$&$+$&$-$&$-$&$-$&$+$&$-$&$-$&$+$&$-$&$-$&$-$&$-$&$+$&$+$&$+$&$-$\\
   &&24&1&$-$&$-$&$+$&$+$&$-$&$-$&$-$&$+$&$+$&$-$&$-$&$+$&$+$&$-$&$+$&$-$&$-$&$+$&$+$&$+$&$+$&$+$&$-$&$-$&$+$&$+$&$+$&$-$&$-$&$+$&$+$&$+$&$-$&$+$&$+$&$-$&$+$&$+$&$+$&$-$&$+$&$+$&$-$&$-$&$+$&$+$&$+$&$-$&$+$&$+$&$+$&$-$\\\hline
$10$&57&28&0&$-$&$0$&$+$&$-$&$-$&$-$&$+$&$+$&$-$&$-$&$-$&$+$&$-$&$+$&$+$&$+$&$-$&$+$&$-$&$-$&$-$&$-$&$+$&$+$&$-$&$-$&$-$&$+$&$+$&$-$&$+$&$+$&$+$&$-$&$-$&$-$&$+$&$+$&$-$&$+$&$-$&$-$&$-$&$+$&$+$&$-$&$+$&$+$&$-$&$+$&$+$&$+$&$+$&$-$&$+$&$+$&$+$&$-$\\
&&28&1&$-$&$-$&$+$&$+$&$-$&$-$&$-$&$+$&$+$&$+$&$-$&$-$&$+$&$-$&$+$&$+$&$-$&$-$&$+$&$-$&$-$&$-$&$-$&$-$&$+$&$+$&$-$&$+$&$+$&$+$&$-$&$-$&$+$&$+$&$+$&$-$&$-$&$-$&$+$&$-$&$+$&$-$&$-$&$-$&$+$&$+$&$-$&$+$&$+$&$-$&$-$&$-$&$+$&$-$&$-$&$+$&$+$&$-$\\\hline
$11$&63&30&0&$-$&$0$&$+$&$-$&$-$&$-$&$+$&$+$&$-$&$+$&$+$&$+$&$-$&$-$&$+$&$+$&$-$&$+$&$-$&$-$&$+$&$+$&$+$&$+$&$-$&$-$&$+$&$+$&$+$&$-$&$-$&$-$&$+$&$-$&$-$&$+$&$+$&$+$&$+$&$-$&$+$&$+$&$-$&$+$&$+$&$+$&$+$&$-$&$+$&$-$&$+$&$+$&$+$&$-$&$+$&$+$&$+$&$+$&$-$&$+$&$+$&$+$&$+$&$-$\\
&&30&1&$-$&$-$&$+$&$+$&$-$&$-$&$-$&$-$&$+$&$-$&$-$&$-$&$+$&$-$&$-$&$+$&$-$&$+$&$+$&$+$&$-$&$+$&$+$&$+$&$+$&$+$&$-$&$-$&$+$&$-$&$-$&$-$&$+$&$+$&$+$&$-$&$-$&$-$&$-$&$+$&$-$&$-$&$+$&$-$&$+$&$+$&$+$&$-$&$+$&$+$&$+$&$-$&$+$&$+$&$-$&$-$&$-$&$+$&$-$&$+$&$+$&$+$&$+$&$-$
 \end{tabular}}
  \label{tb:sturml}
 \end{table*}
\subsection{Proof of Lemma \ref{lemma:II}}
\label{sec:154724_18Jan14}
We first claim that for $z \in(0,1)$,
\begin{align}
&\text{If } \frac{a\dl+b}{a\dl+b+1}\in(0,1), \text{ then}\\
& q(z):=z^{a\dl+b}(1-z) \le \frac{1}{a\dl+b+1},\label{152517_18Jan14} \\
&\text{If }\frac{a\dl+b}{(a+2)\dl+b-2}\in(0,1), \text{ then}\\
& r(z):=z^{a\dl+b}(1-z^{\dl-1})^{2} \le \Bigl(\frac{2\dl-2}{(a+2)\dl+b-2}\Bigr)^2\label{152506_18Jan14}
\end{align} 
Differentiating $q(z)$ gives
\begin{align}
 \frac{\d q(z)}{\d z}
 &=z^{a\dl+b-1}(a\dl+b-(a\dl+b+1)z).
 \end{align}
Since $\frac{a\dl+b}{a\dl+b+1}\in(0,1)$, we see that $z=\frac{a\dl+b}{a\dl+b+1}$ gives the maximum value of $q(z)$. 
 \begin{align}
   q(z)\le \Bigl(\frac{a\dl+b}{a\dl+b+1}\Bigr)^{a\dl+b}\frac{1}{a\dl+b+1}<\frac{1}{a\dl+b+1}.
 \end{align}
Differentiating $r(z)$  gives 
 \begin{align}
& \frac{\d r(z)}{\d z}
 =z^{a\dl+b-1}(1-z^{l-1})((a\dl+b)-((a+2)\dl+b-2)z^{l-1}).
 \end{align}
Since $\frac{a\dl+b}{(a+2)\dl+b-2}\in(0,1)$, we see that 
$z=\bigl(\frac{a\dl+b}{(a+2)\dl+b-2}\bigr)^{\frac{1}{\dl-1}}$ gives the maximum value of $r(z)$. 
Thus, next inequality holds.
 \begin{equation}
  \begin{split}
   r(z)&\le 
\Bigl(\frac{a\dl+b}{(a+2)\dl+b-2}\Bigr)^{\frac{a\dl+b}{\dl-1}}\Bigl(\frac{2\dl-2}{(a+2)\dl+b-2}\Bigr)^{2}\\
   &<\Bigl(\frac{2\dl-2}{(a+2)\dl+b-2}\Bigr)^{2}.
  \end{split}
 \end{equation}

In (a), we eliminate negative terms except for $-\dl^3$. 
Next, in (b), we apply \eqref{152517_18Jan14} and  \eqref{152506_18Jan14} to each term of \eqref{eq:I} by
using $\dl\ge 165.$
We obtain an upper bound of $I(z)$ for $z\in(0,1)$ as follows. 
\begin{align}
 I(z)\stackrel{\mathrm{(a)}}{<}&-\dl^3+27\sum_{i=0}^{\dl-2}z^{3\dl-2+i}(1 - z^{\dl-1})\\
&+108\dl^2z^{-3 + 3\dl} (1 - z^{\dl-1})^2 \\
 &+ 9lz^{-4 + l}(1 - z^{\dl-1})^2(3z^2+ 16 z^{2l}+ 8z^{2 + \dl}) \\
 &+\dl^3\{ (1 - z)z^{-9 + \dl}[  56z^{1 + 5\dl}   +8z^{3 + 3\dl}(13 + 22z) \\
 &+ z^{5 + \dl}(41 + 73z)]\}\\
 \stackrel{\mathrm{(b)}}{<}&-l^3+27\sum_{i=0}^{\dl-2}[1]+108\dl^{2}\bigl(\frac{2\dl-2}{5\dl-5}\bigr)^{2}\\
&+9\dl \Bigl(3\bigl(\frac{2\dl-2}{3\dl-4}\bigr)^{2}
 +16\bigl(\frac{2\dl-2}{5\dl-6}\bigr)^{2}+8\bigl(\frac{2\dl-2}{4\dl-4}\bigr)^{2}\Bigr)\\
 &+\dl^{3}\Bigl(\frac{56}{6\dl-7}+\frac{176}{4\dl-4}+\frac{104}{4\dl-5}+\frac{41}{2\dl-3}+\frac{73}{2\dl-2}\Bigr)\\
  \stackrel{\mathrm{(c)}}{<}&-\dl^{3}+27(l-1)+\frac{432\dl^{2}}{25}+9\dl\Bigl(3\frac{5}{9}+16\frac{1}{5}+8\frac{1}{4}\Bigr)\\
 &+5\dl^{3}\Bigl(\frac{59}{29\dl }+\frac{176}{19\dl}+\frac{104}{19\dl }+\frac{73}{9\dl }+\frac{41}{9\dl }\Bigr)\\
 <&-\dl^{3}+\frac{6775346}{41325}\dl^{2}+\frac{444}{5}\dl=:\overline{I}(l).
\end{align}
We used next inequalities valid for $\dl \ge165$ in (c). 
 \begin{eqnarray*}
\Bigl( \frac{2\dl-2}{3\dl-4}\Bigr)^{2} \le  \frac{5}{9} ,&&  \Bigl(\frac{2\dl-2}{5\dl-6}\Bigr)^{2} \le \frac{1}{5},\\
 6\dl-7 \ge \frac{29\dl}{5},&&  4\dl-4 \ge \frac{19\dl}{5},\\
 4\dl-5 \ge \frac{19\dl}{5},&& 2\dl-3 \ge \frac{9\dl}{5},\\
 2\dl-2 \ge \frac{9\dl}{5}. &&
 \end{eqnarray*}
The roots of $\overline{I}(l)=0$ are $0$ and $\frac{3387673\pm\sqrt{11627977054429}}{41325}\simeq -0.53984,+164.49. $
Thus, we conclude that  for $\dl \ge 165$ and $z\in(0,1)$, $I(z)<\overline{I}(\dl)<0$. \qed
 \section{Conclusion and Future Work}
In this paper, we proved that $(l,3,3)$ SC-MN codes with $l \ge 3$ achieve capacity on the BEC under BP decoding for sufficiently large $L$ and $w$. 
This codes do not have bit nodes of degree two and have low error floors. 
We proved that the potential threshold  and Shannon limit of ($\dl,\dr=3,\dg=3$) MN codes  on the BEC are the same.

\bibliographystyle{IEEEtran}
\bibliography{IEEEabrv,../../kenta_bib}

\begin{thebibliography}{1}
\providecommand{\url}[1]{#1}
\csname url@rmstyle\endcsname
\providecommand{\newblock}{\relax}
\providecommand{\bibinfo}[2]{#2}
\providecommand\BIBentrySTDinterwordspacing{\spaceskip=0pt\relax}
\providecommand\BIBentryALTinterwordstretchfactor{4}
\providecommand\BIBentryALTinterwordspacing{\spaceskip=\fontdimen2\font plus
\BIBentryALTinterwordstretchfactor\fontdimen3\font minus
  \fontdimen4\font\relax}
\providecommand\BIBforeignlanguage[2]{{%
\expandafter\ifx\csname l@#1\endcsname\relax
\typeout{** WARNING: IEEEtran.bst: No hyphenation pattern has been}%
\typeout{** loaded for the language `#1'. Using the pattern for}%
\typeout{** the default language instead.}%
\else
\language=\csname l@#1\endcsname
\fi
#2}}

\bibitem{zigangirov99}
A.~J. Felstr{\"o}m and K.~S. Zigangirov, ``Time-varying periodic convolutional
  codes with low-density parity-check matrix,'' \emph{{IEEE} Trans. Inf.
  Theory}, vol.~45, no.~6, pp. 2181--2191, June 1999.

\bibitem{lentmaier_II}
M.~Lentmaier, D.~V. Truhachev, and K.~S. Zigangirov, ``To the theory of
  low-density convolutional codes. {II},'' \emph{Probl.~ Inf.~ Transm.~},
  no.~4, pp. 288--306, 2001.

\bibitem{5695130}
S.~Kudekar, T.~Richardson, and R.~Urbanke, ``Threshold saturation via spatial
  coupling: Why convolutional {LDPC} ensembles perform so well over the
  {BEC},'' \emph{{IEEE} Trans. Inf. Theory}, vol.~57, no.~2, pp. 803--834, Feb.
  2011.

\bibitem{6589171}
------, ``Spatially coupled ensembles universally achieve capacity under belief
  propagation,'' \emph{{IEEE} Trans. Inf. Theory}, vol.~59, no.~12, pp.
  7761--7813, 2013.

\bibitem{HSU_MN_IEICE}
K.~Kasai and K.~Sakaniwa, ``Spatially-coupled {M}ac{K}ay-{N}eal codes and
  {H}su-{A}nastasopoulos codes,'' \emph{IEICE Trans. Fundamentals}, vol. E94-A,
  no.~11, pp. 2161--2168, Nov. 2011.

\bibitem{ISIT_OJKP}
N.~Obata, Y.-Y. Jian, K.~Kasai, and H.~D. Pfister, ``Spatially-coupled
  multi-edge type {LDPC} codes with bounded degrees that achieve capacity on
  the {BEC} under {BP} decoding,'' in \emph{Proc. 2013 {IEEE} Int. Symp. Inf.
  Theory (ISIT)}, July 2013, pp. 2433--2437.

\bibitem{simple_proof_BEC_itw}
A.~Yedla, Y.-Y. Jian, P.~Nguyen, and H.~Pfister, ``A simple proof of threshold
  saturation for coupled scalar recursions,'' in \emph{Turbo Codes and
  Iterative Information Processing (ISTC), 2012 7th International Symposium
  on}, 2012, pp. 51--55.

\bibitem{gautschi2011numerical}
W.~Gautschi, \emph{Numerical Analysis}, ser. SpringerLink : B{\"u}cher.\hskip
  1em plus 0.5em minus 0.4em\relax Springer Science+Business Media, LLC, 2011.

\end{thebibliography}
\appendix[Sturm's Theorem]
\begin{theorem}[\cite{gautschi2011numerical}]
\label{theorem:sturm}
For a polynomial $f(x)$ over $\R$, we define Sturm sequences $f_i(x)\ (i=0,\dotsc , m)$ 
as $f(x)$, $f'(x)$ and polynomials 
obtained by applying Euclid's algorithm to $f(x)$ and $f'(x)$.
 \begin{eqnarray*}
f_0(x)&=&f(x),\\ 
f_1(x)&=&f'(x),\\
 f_{n-1}(x)&=&q_n(x)f_n(x)-f_{n+1}(x)\quad (n=1,\dotsc,m-1),\\
f_{m-1}(x)&=&q_m(x)f_m(x). 
\end{eqnarray*}
For real number $c$, let $V(c)$ be the number of sign changes in $f_0(c), f_1(c), \dotsc, f_m(c)$. If neither $a\in \R$ nor $b\in \R$ is a multiple root of $f(x)=0$, 
then the number of distinct roots of $f(x)$ in $(a,b]$ is $V(a)-V(b)$.
\end{theorem}
\end{document}